\documentclass[twocolumn,showpacs,preprintnumbers,amsmath,amssymb]{revtex4}

\usepackage{graphicx}
\usepackage{dcolumn}
\usepackage{bm}

\begin{document}

\title{Neutron spectroscopic factors of $^7$Li and astrophysical $^6$Li(n,$\gamma)^7$Li reaction rates
\footnote{Supported by the the National Basic Research Programme of
China under Grant No. 2007CB815003, the National Natural Science
Foundation of China under Grant Nos.10675173, 10720101076, 10735100
and 10975193. } }
\author{SU Jun}
\email{junsu@iris.ciae.ac.cn}
\author{LI Zhi-Hong}
\author{GUO Bing}
\author{\\BAI Xi-Xiang}
\author{LI Zhi-Chang}
\author{LIU Jian-Cheng}
\author{WANG You-Bao}
\author{LIAN Gang}
\author{ZENG Sheng}
\author{\\WANG Bao-Xiang}
\author{YAN Sheng-Quan}
\author{LI Yun-Ju}
\author{LI Er-Tao}
\author{FAN Qi-Wen}
\author{LIU Wei-Ping}

\affiliation{China Institute of Atomic Energy, P. O. Box 275(46),
Beijing 102413, P. R. China}%

\begin{abstract}
Angular distributions of the  $^7$Li($^6$Li,$^6$Li)$^7$Li elastic
scattering and the $^7$Li($^6$Li,$^7$Li$_{g.s.})^6$Li,
$^7$Li($^6$Li,$^7$Li$_{0.48})^6$Li transfer reactions at
E$_{c.m.}$=23.7  MeV were measured with the Q3D magnetic
spectrograph. The optical potential of $^6$Li+$^7$Li was obtained by
fitting the elastic scattering differential cross sections. Based on
the distorted wave Born approximation (DWBA) analysis, spectroscopic
factors of $^7$Li$=^6$Li$\otimes$n were determined to be
$0.73\pm0.05$ and $0.90\pm0.09$ for the ground and first exited
states in $^7$Li, respectively. Using the spectroscopic factors, the
cross sections of the $^6$Li(n,$\gamma_{0,1})^7$Li direct neutron
capture reactions and the astrophysical $^6$Li(n,$\gamma)^7$Li
reaction rates were derived.
\end{abstract}

\maketitle

Recently, lithium isotopes have attracted an intense interest
because the abundance of both $^6$Li and $^7$Li from big bang
nucleosynthesis (BBN) is one of puzzles in nuclear astrophysics.
According to the baryon density determined by Wilkinson microwave
anisotropy probe (WMAP)\cite{Dun09}, the primary abundances for
$^6$Li and $^7$Li predicted by standard BBN (SBBN) model deviate
clearly from the observations of the metal-poor halo
stars\cite{Cay08}, where the lithium abundances exhibit a "plateau"
behavior\cite{Spi82}. Several investigations for both astrophysical
observation and nucleosynthesis calculation have been attempted to
explain the large discrepancies, but none of them has been
successful up to now. In addition, due to the difference between the
depletion speeds of $^6$Li and $^7$Li in stars, the $^6$Li/$^7$Li
ratio could stand for a measure of the time scale for stellar
evolution. In the above scenario, $^6$Li(n,$\gamma)^7$Li is believed
to be one of the important reactions in the SBBN
network\cite{Mal89,Nol97}, its reaction rates would affect the
abundances of both $^6$Li and $^7$Li.

The cross sections of $^6$Li(n,$\gamma)^7$Li at astrophysically
relevant energies are most likely dominated by the E1 transitions
into the ground and first exited states in $^7$Li. To date, only one
direct measurement of the $^6$Li(n,$\gamma)^7$Li cross sections at
stellar energies has been performed\cite{Tos00}. The cross sections
can also be calculated by the direct capture model with the
spectroscopic factors of $^7$Li$=^6$Li$\otimes$n extracted from the
neutron transfer reactions. In the previous studies, the
spectroscopic factors were mostly derived by $^6$Li(d, p)$^7$Li and
$^7$Li(p, d)$^6$Li reactions\cite{Sch67,Li69,Tow69,Fag76,Tsa05}, the
results are correlative with the neutron spectroscopic factor of
deuteron. Thus, it is highly desired to extract the spectroscopic
factors through a self-contained reaction without third-participant,
which can provide an independent verification.

The elastic-transfer reaction is a good tool to extract the single
nucleon spectroscopic factors, which has been used to determine the
spectroscopic factor of $^9$Be$=^8$Li$\otimes$p with
$^9$Be($^8$Li,$^9$Be)$^8$Li reaction\cite{Cam08}. In such an
approach, the events from elastic scattering and elastic-transfer
reaction can not be distinguished, but the theoretical calculation
shows that their mutual contributions at the respective forward
angles are negligibly small. As a result, the effective differential
cross sections at forward angles of the two processes can be
obtained respectively. The obvious advantages of this approach are
(i) the elastic scattering and elastic-transfer reaction have the
same entrance and exit channels, the spectroscopic factor can be
derived by one set of optical potential deduced from the elastic
scattering and (ii) the elastic-transfer reaction does not involve
third participant in the entrance and exit channels, and thus
conduces to the reduction of experimental result uncertainty.

For the above reasons, we chose the $^7$Li($^6$Li,$^7$Li)$^6$Li
elastic-transfer reaction to extract the spectroscopic factors of
the $^7$Li$=^6$Li$\otimes$n. This reaction has been measured in 1998
at E$_{lab}$=9-40 MeV\cite{Pot98}, unfortunately the minimum angle
reached in that experiment was about 30$^{\circ}$ in the center of
mass frame (for transfer process), that was not suitable to derive
the spectroscopic factor. In present work, we have measured the
angular distributions of the $^7$Li($^6$Li,$^6$Li)$^7$Li elastic
scattering and $^7$Li($^6$Li,$^7$Li$_{g.s.})^6$Li,
$^7$Li($^6$Li,$^7$Li$_{0.48})^6$Li transfer reactions at
E$_{c.m.}$=23.7 MeV. The neutron spectroscopic factors for the
ground and first exited states in $^7$Li were determined by
comparing the experimental results with the distorted-wave Born
approximation (DWBA) calculations, and then used to calculate the
cross sections and astrophysical rates of $^6$Li(n,$\gamma)^7$Li
direct capture reaction.

The experiment was carried out at the Beijing HI-13 tandem
accelerator. A 44 MeV $^6$Li beam in intensity of about 100 pnA
impinged on the natural LiF target in thickness of 530 $\mu g/cm^2$,
which was evaporated on a 50 $\mu g/cm^2$ carbon foil. The beam was
collected by a Faraday cup behind the target for counting the number
of $^6$Li. The Faraday cup covered an angle range of $\pm6^{\circ}$
and confined the attainable minimum angle in the measurement. The
reaction products were focused and separated by Q3D magnetic
spectrograph. The accepted solid angle of Q3D was set to be  0.23
mSr for a better angular resolution. A two-dimensional position
sensitive silicon detector (PSSD) was set at the focal plane of Q3D,
the X-Y information from PSSD enabled the products emitted into the
accepted solid angle of Q3D to be fully recorded, and the
corresponding energy signals were used to remove the impurities with
the same magnetic rigidity. The absolute differential cross sections
were determined by normalizing the measurements of the
$^7$Li($^6$Li,$^6$Li)$^7$Li elastic scattering and the
$^7$Li($^6$Li,$^7$Li$_{g.s.})^6$Li,
$^7$Li($^6$Li,$^7$Li$_{0.48})^6$Li transfer reactions to the elastic
scattering of $^6$Li on the gold target at
$\theta_{lab}=25^{\circ}$.

\begin{figure}[htcp]
\includegraphics[height=6 cm]{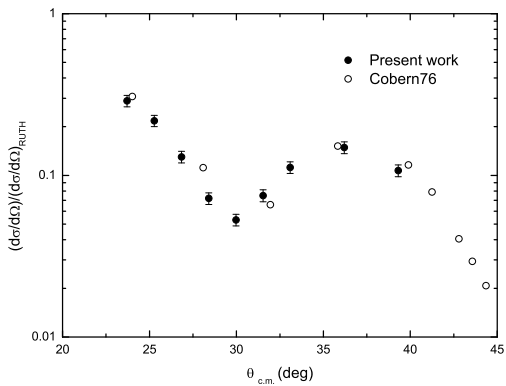}
\caption{\label{fig:li7c12} Comparison of the
$^{12}$C($^7$Li,$^7$Li)$^{12}$C angular distributions at
E$_{lab}$=36 MeV, the solid and open circles denote the data
obtained in the present experiment and an earlier work\cite{Cob76},
respectively.}
\end{figure}

The experiment setup was tested beforehand by measuring the angular
distribution of $^{12}$C($^7$Li,$^7$Li)$^{12}$C elastic scattering
at E$_{lab}$=36 MeV. As shown in the Fig.~\ref{fig:li7c12}, our
result is in fair agreement with that reported in Ref.\cite{Cob76},
indicating a reliable overall performance of our setup and data
analysis procedure.

In the measurements of $^7$Li($^6$Li,$^6$Li)$^7$Li elastic
scattering and $^7$Li($^6$Li,$^7$Li$_{g.s.})^6$Li,
$^7$Li($^6$Li,$^7$Li$_{0.48})^6$Li transfer reactions, the magnetic
fields of Q3D were set to focus $^6$Li and $^7$Li, respectively. The
elastic scattering and transfer processes were measured in the
angular ranges of $7^{\circ}\leq\theta_{lab}\leq 30^{\circ}$ and
$7^{\circ}\leq\theta_{lab}\leq 17^{\circ}$ in steps of $1^{\circ}$,
respectively.

\begin{figure}[htcp]
\includegraphics[height=6 cm]{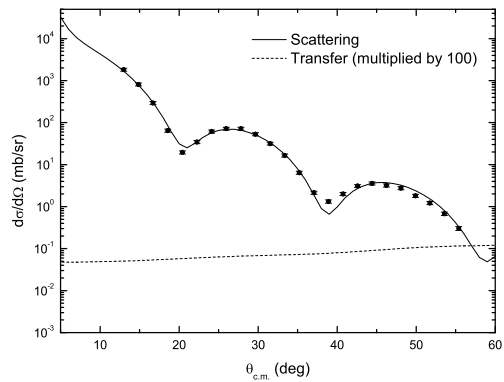}
\caption{\label{fig:elastic} Differential cross sections for
$^7$Li($^6$Li,$^6$Li)$^7$Li elastic scattering at E$_{c.m.}$=23.7
MeV together with the optical model calculation. The dashed line
denotes the contribution of transfer processes (multiplied by a
factor of 100). }
\end{figure}

The differential cross sections for elastic scattering are shown in
the Fig.~\ref{fig:elastic}, with uncertainties from the errors of
statistics and target thickness. The angular distribution of
$^7$Li($^6$Li,$^6$Li)$^7$Li elastic scattering was analyzed by
employing the code PTOLEMY\cite{Mac} and optical model using real
and imaginary potentials with Woods-Saxon form . The optimized
potential parameters obtained by fitting the experimental data are
listed in the Table ~\ref{tab:tab1}. The fitting results are shown
in Fig.~\ref{fig:elastic} with the experimental data.
Fig.~\ref{fig:elastic} also exhibits the contribution of transfer
processes, which is less than 1\% in the experimental angular range.

\begin{table}
\caption{\label{tab:tab1} Optical potential parameters of
$^6$Li+$^7$Li at E$_{c.m.}$=23.7 MeV. The depths and geometrical
parameters are in MeV and fm, respectively.\\}
\begin{tabular*}{8.5 cm}{@{\extracolsep{\fill}}cccccccc}
 \hline
$U_V$& $r_R$& $a_R$& $W_V$&$r_I$&$a_I$&$r_c$&$\chi^2/point$\\
\hline 73.4&0.5&0.92&32.6&1.0&0.8&1.25&4.0\\

 \hline
\end{tabular*}
\end{table}

\begin{figure}[htcp]
\includegraphics[height=7 cm]{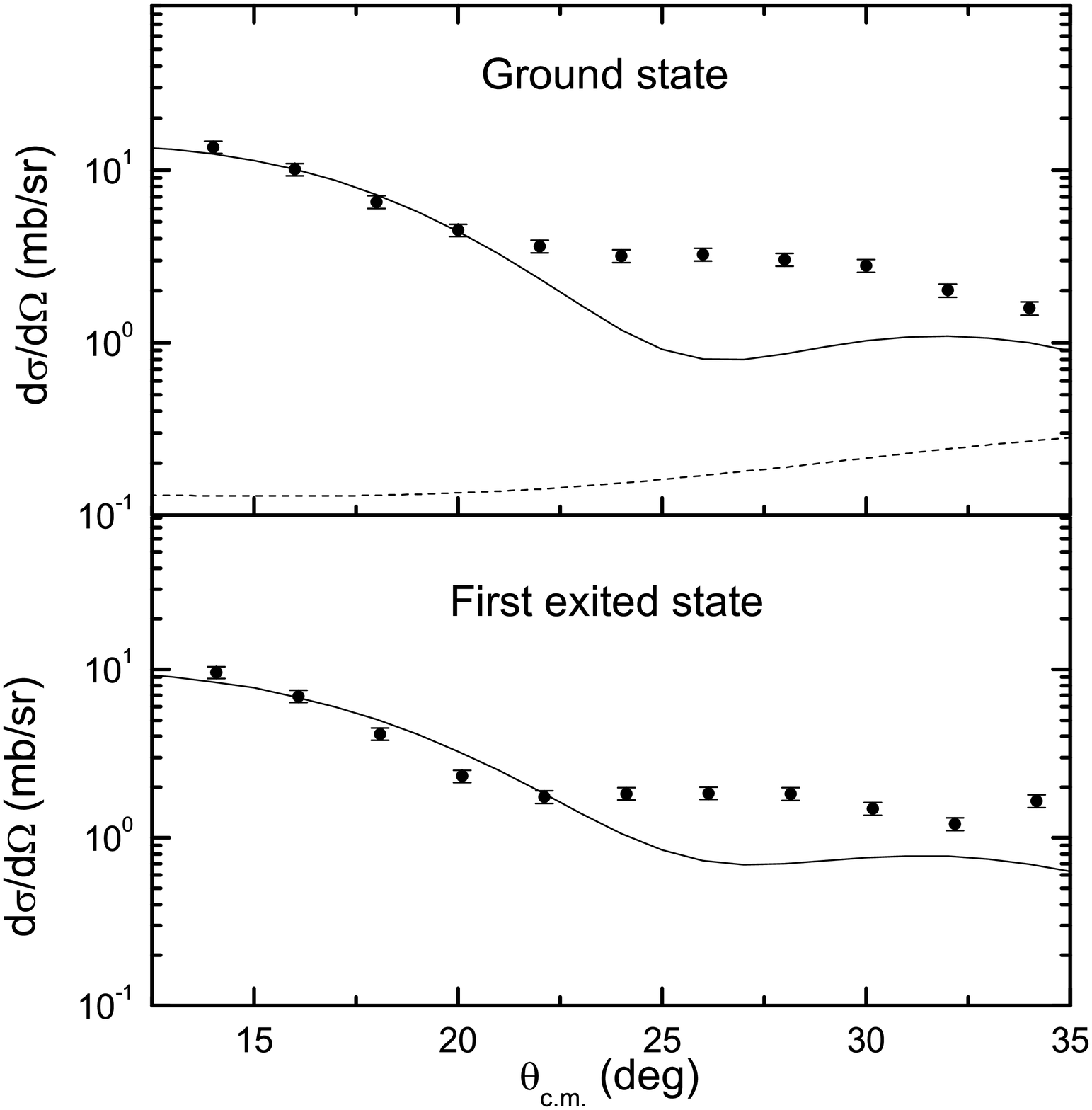}
\caption{\label{fig:transfer} Angular distribution of $^7$Li($^6$Li,
$^7$Li$_{g.s.})^6$Li and $^7$Li($^6$Li, $^7$Li$_{0.48})^6$Li at
E$_{c.m.}$=23.7 MeV together with the DWBA calculations normalized
to the first three data at forward angles. The dash line denotes the
contribution of elastic scattering (multiplied by a factor of
1000).}
\end{figure}

The angular distributions of $^7$Li($^6$Li, $^7$Li$_{g.s.})^6$Li and
$^7$Li($^6$Li, $^7$Li$_{0.48})^6$Li transfer processes are shown in
Fig.~\ref{fig:transfer}, which were also analyzed with the code
PTOLEMY. As can clearly be seen from Fig.~\ref{fig:transfer}, the
contribution of elastic scattering is negligible small. In the
calculation, we utilized the $^6$Li+$^7$Li optical potential
parameters listed in Table~\ref{tab:tab1} for both the entrance and
exit channels. For the bound states, a Woods-Saxon potential with
the standard geometrical parameters $r_0$=1.25 fm and a=0.65 fm was
adopted and the depths were adjusted to reproduce the neutron
binding energies of $^7$Li.

The spectroscopic factor $^7$Li=$^6$Li$\otimes$n, denoted as
S$_{^7Li}$, can be derived by normalizing the DWBA calculations to
the experimental data according to the expression
\begin{equation}
(\frac{d\sigma}{d\Omega})_{EXP}=S_{^7Li}^2(\frac{d\sigma}{d\Omega})_{DWBA}.
\end{equation}

Generally, the experimental data in the first peak of angular
distributions at the forward angles are suitable for extracting the
spectroscopic factor\cite{Liu04} because the differential cross
sections of other angles are more sensitive to some high-order
processes. In our calculation, only the first three data at the
forward angles were used to extract the spectroscopic factors. For
the $^7$Li($^6$Li, $^7$Li$_{g.s.})^6$Li channel, both of the neutron
transfers to the 1p$_{3/2}$ and 1p$_{1/2}$ orbits of $^7$Li were
taken into account, consequently the result was the coherent sum of
their contributions. Since the 1p$_{3/2}$ and 1p$_{1/2}$ components
can not be distinguished experimentally, their ratio was taken to be
1.5 based on the calculation\cite{Coh67}, then the spectroscopic
factor of $^7$Li$_{g.s.}=^6$Li$\otimes$n was determined to be
$0.73\pm0.05$, the uncertainty results from the measurement,
scattering potential fitting and the divergence of the values
derived from three different angles. For the first exited state in
$^7$Li, only the neutron transfer to the 1p$_{3/2}$ orbit was
considered because the contribution of 1p$_{1/2}$ orbit is merely
about 4\%\cite{Coh67}, then the spectroscopic factor of
$^7$Li$_{0.48}=^6$Li$\otimes$n was derived to be $0.90\pm0.09$. The
spectroscopic factors obtained in several theoretical and
experimental investigations are listed in Table~\ref{tab:tab2}, our
results agree well with the theoretical
calculation\cite{Coh67,Bar66,Var69,Kum74} and the experimental
data\cite{Li69,Tow69}.

\begin{table}[h]
\caption{\label{tab:tab2} The theoretical and experimental neutron
spectroscopic factors for the ground and first exited states in
$^7$Li.\\}
\begin{tabular*}{8.5 cm}{@{\extracolsep{\fill}}cccc}
\hline\hline
$S_{^7Li_{g.s.}}$& $S_{^7Li_{0.48}}$& Experiments or theory& Reference\\
\hline
0.72&0.89&theory&\cite{Coh67}\\
0.80&0.98&theory&\cite{Bar66}\\
0.79&0.97&theory&\cite{Var69}\\
0.77&1.07&theory&\cite{Kum74}\\
0.90&1.15&$^6$Li(d,p)&\cite{Sch67}\\
0.71&    &$^7$Li(p,d)&\cite{Li69}\\
$0.72\pm0.1$&    &$^7$Li(p,d)&\cite{Tow69}\\
0.87&    &$^7$Li(p,d)&\cite{Fag76}\\
$1.85\pm0.37$&&$^6$Li(d,p)&       \cite{Tsa05}\\
$0.73\pm0.05$&$0.90\pm0.09$&$^7$Li($^6$Li,$^7$Li)&present work\\
 \hline\hline
\end{tabular*}
\end{table}

At low energies of astrophysical interest, the
$^6$Li(n,$\gamma)^7$Li direct capture cross sections are dominated
by the E1 transition from incoming s-wave to the bound state, the
contribution of d-wave neutrons is negligible. In the computation we
used the code RADCAP\cite{Ber03} and Woods-Saxon form potential with
the standard geometrical parameters. The potential depths of the
$^7$Li bound states were adjusted to reproduce the neutron binding
energies of its ground and first excited states, respectively. For
the scattering potential, the depth can be fixed by fitting the well
measured $^6$Li(p,$\gamma)^7$Be cross sections at low
energies\cite{Swi79} because the difference between n+$^6$Li and
p+$^6$Li potentials is only a Coulomb modification of depth, which
is given by $\Delta V=0.4Z/A^{1/3}$\cite{Dav83}.

\begin{figure}
\includegraphics[height=5.5 cm]{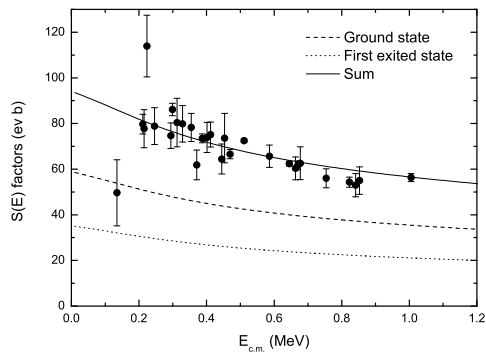}
\caption{\label{fig:li6pg} Astrophysical S(E) factors of the
$^6$Li(p,$\gamma)^7$Be reaction. The experimental data are taken
from Ref.\cite{Swi79}.}
\end{figure}
The $^6$Li(p,$\gamma)^7$Be cross sections were also calculated  by
the same code. The neutron spectroscopic factors of $^7$Li derived
above were correspondingly taken as the proton spectroscopic factors
for the ground and first exited states in $^7$Be owing to the charge
symmetry. Here, the depth of scattering potential was constrained to
be $41.3\pm2.0$ MeV in reproducing the experimental data. The
calculated astrophysical $^6$Li(p,$\gamma)^7$Be S(E) factors are
shown in Fig.~\ref{fig:li6pg} together with the experimental
results. In addition to a good agreement between the calculated and
measured total S(E) factors, our calculation also indicates that the
contributions of ground and first exited states are about 63\% and
37\%, respectively, which are very close to the experimental values
61\% and 39\%\cite{Swi79}.

\begin{figure}
\includegraphics[height=5.5 cm]{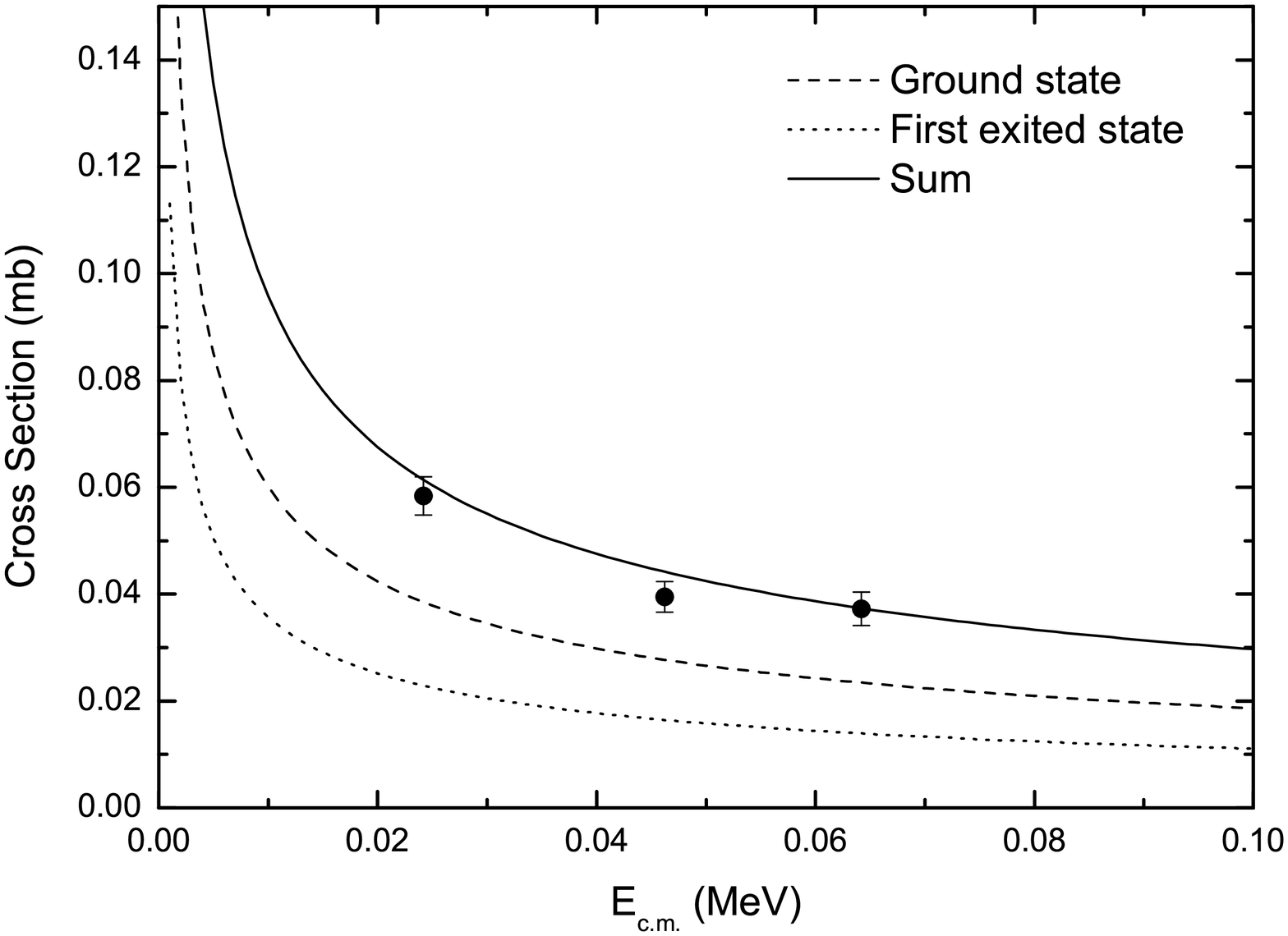}
\caption{\label{fig:li6ng} Cross sections of the
$^6$Li(n,$\gamma)^7$Li reaction. The experimental data are taken
from Ref.\cite{Tos00}.}
\end{figure}
Considering the Coulomb modification, the scattering potential depth
of n+$^6$Li was chosen to be $40.6\pm2.0$ MeV. Then the
$^6$Li(n,$\gamma)^7$Li$_{g.s.}$ and $^6$Li(n,$\gamma)^7$Li$_{0.48}$
cross sections were calculated using the spectroscopic factors and
optical potentials extracted above, and compared with the
experimental data\cite{Tos00}, as shown in Fig.~\ref{fig:li6ng}. Our
result is consistent with the direct measurement, showing an
approximate 1/v behavior, so that the reaction rates are almost
constant at energies of astrophysical interest.

The astrophysical $^6$Li(n,$\gamma)^7$Li direct capture reaction
rate was then calculated by the expression\cite{Ang99}
\begin{eqnarray}\label{equ:rates}
N_A\langle \sigma v\rangle
=3.73\times10^{10}A^{-\frac{1}{2}}T^{-\frac{3}{2}}_9\int\limits_{0}^{\infty}\sigma
E exp(\frac{-11.6E}{T_9})dE,
\end{eqnarray}
where A is the reduced mass in amu, T$_9$ is the temperature in
units of 10$^9$K. E, $\sigma$ and reaction rate are given in MeV,
barns and cm$^3$mol$^{-1}$s$^{-1}$, respectively. The reaction rate
was found to be $(8.5\pm1.7)\times10^{3}$ cm$^3$mol$^{-1}$s$^{-1}$,
the error results from the uncertainties of spectroscopic factors
and scattering potential depth.

Summarizing, the measurements of differential cross sections for the
$^7$Li($^6$Li,$^6$Li)$^7$Li elastic scattering and
$^7$Li($^6$Li,$^7$Li$_{g.s.})^6$Li,
$^7$Li($^6$Li,$^7$Li$_{0.48})^6$Li transfer reactions have been
carried out at E$_{c.m.}$=23.7 MeV, in which the angular
distributions for transfer processes were obtained in the range of
$14^{\circ}\lesssim\theta_{c.m. }\lesssim 34^{\circ}$ for the first
time. By using the optical potential of $^6$Li+$^7$Li extracted from
the elastic scattering, the spectroscopic factors of
$^7$Li$=^6$Li$\otimes$n were deduced with DWBA analysis, the results
are in agreement with those reported
previously\cite{Li69,Tow69,Coh67,Bar66,Var69,Kum74}. Then the
$^6$Li(n,$\gamma)^7$Li direct capture cross sections have been
derived and compared with the direct measurement data. The
astrophysical reaction rate was found to be higher by a factor of
1.7 than the value adopted in previous reaction network
calculations\cite{Mal89,Nol97}.

\end{document}